\documentclass[twocolumn,twoside,slac_two]{revtex4}
\usepackage{graphicx}
\usepackage{fancyhdr}
\begin{document}

\title{Charmonium Spectroscopy Below Open Flavor Threshold}

%

\author{Z. Metreveli}
\affiliation{Department of Physics, Northwestern University, Evanston, IL 60208, USA}

\begin{abstract}
Latest experimental results in the charmonium spectroscopy below $D\bar{D}$ 
breakup threshold are reviewed.
\end{abstract}

\maketitle

\thispagestyle{fancy}


\section{Introduction}

Charm quark has large mass ($\sim$1.5 GeV) compared to the masses of
{\it u, d, s} quarks. Velocity of the charm quarks in hadrons is not
too relativistic ({\it (v/c)}$^{2}\sim$0.2). Strong coupling constant
$\alpha_{s}(m_{c})$ is small ($\sim$0.3). Therefore charmonium 
spectroscopy is a good testing ground for the theories of strong interactions:
quantum cromodynamics (QCD) in both perturbative and nonperturbative regimes,
QCD inspired purely phenomenological potential models, nonrelativistic
QCD (NRQCD) and lattice QCD.

\begin{figure}[h]
\centering
\includegraphics[width=60mm]{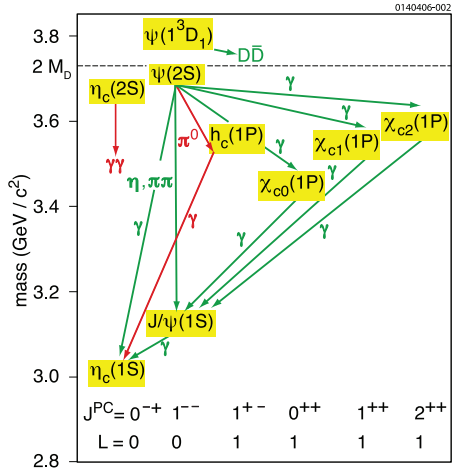}
\caption{Spectra of the states of charmonium below $D\bar{D}$ breakup 
threshold.} \label{plot1}
\end{figure}

There are 8 bound states of charmonium below the $D\bar{D}$ breakup
threshold (Fig.~\ref{plot1}). These are spin triplets 
$J/\psi$($1^{3}S_{1}$), $\psi(2S)$($2^{3}S_{1}$), 
$\chi_{c0,1,2}$($1^{3}P_{0,1,2}$) and spin singlets 
$\eta_{c}$($1^{1}S_{0}$), $\eta_{c}(2S)$($2^{1}S_{0}$),
$h_{c}$($1^{1}P_{1}$). Only $J/\psi$ and $\psi(2S)$ can be produced
directly in $e^{+}e^{-}$ annihilation. A lot is known about these triplet
states. Spin singlet states population via radiative transitions from
the vector states is either very weak (M1 transitions for $\eta_{c}(1S)$,
$\eta_{c}(2S)$), or {\it C}-forbidden ($h_{c}(1^{1}P_{1})$). Accordingly,
little is known about these singlet states.

The status of charmonium states below $D\bar{D}$ breakup threshold is
summarized in Table~\ref{charmstat}. The masses and widths from PDG 2007,
as well as a number of measured decay channels from PDG 2002, 2004 and 2007
are presented separately for {\it spin-triplet} and {\it spin-singlet} 
states. 

\begin{table}[h]
\begin{center}
\caption{Status of charmonium states.}
\begin{tabular}{|c|c|c|c|c|c|}
\hline
& Mass & Width & \multicolumn{3}{c|}{Number of Decays} \\
& (MeV) & (MeV) & \multicolumn{3}{c|}{PDG} \\
& PDG 2007 & PDG 2007 & 2002 & 2004 & 2007$^{\star}$ \\ \hline
\multicolumn{6}{|c|}{Spin Triplets} \\ \hline
$J/\psi$ & 3096.92$\pm$0.01 & 93.4$\pm$2.1 (keV) & 134 & 135 & 162 \\ \hline
$\psi(2S)$ & 3686.09$\pm$0.03 & 327$\pm$11 (keV) & 51 & 62 & 115 \\ \hline
$\chi_{c0}$ & 3414.75$\pm$0.35 & 10.4$\pm$0.7 & 17 & 17 & 51 \\ \hline
$\chi_{c1}$ & 3510.66$\pm$0.07 & 0.89$\pm$0.05 & 12 & 13 & 35 \\ \hline
$\chi_{c2}$ & 3556.20$\pm$0.09 & 2.05$\pm$0.12 & 18 & 19 & 37 \\ \hline
\multicolumn{6}{|c|}{Spin Singlets} \\ \hline
$\eta_{c}(1S)$ & 2979.8$\pm$1.2 & 26.5$\pm$3.5 & 20 & 21 & 31 \\ \hline
$\eta_{c}(2S)$ & 3637$\pm$4 & 14$\pm$7 & 3 & 4 & 4 \\ \hline
$h_{c}$ & 3525.93$\pm$0.27 & $<$1 & 3 & 3 & 4 \\ \hline
\end{tabular}
\label{charmstat}
\end{center}
\end{table}

It is obvious from Table~\ref{charmstat} that the parameters
of {\it spin-triplet} states are measured with precision, and the
number of measured decay channels is large (notice the marked improvements
after 2004). This is not valid for {\it spin-singlet} states. A lot
remains to be done for precision measurements of their parameters and
decay channels.

Some new and recent experimental developments on charmonium spectroscopy
below open flavor threshold will be reviewed.

\section{Observation of $\eta_{c}(2S)$}

It is important to identify the spin-singlet states in order to determine
the hyperfine, or {\it spin-spin} interaction, which is responsible for
singlet-triplet splitting of $q\bar{q}$ states. Identification of 
$\eta_{c}(2S)$ is important to know the possible variation of spin-spin 
interaction from Coulombic ($J/\psi$, $\eta_{c}(1S)$) to Confinement
($\psi(2S)$, $\eta_{c}(2S)$) regions of the $q\bar{q}$ interaction.
Most potential model calculations predicted $M(\eta_{c}(2S)$=3594-3626 (MeV).

Prior to 2002 there were several unsuccessful attempts to identify
$\eta_{c}(2S)$ in $p\bar{p}$, $\gamma \gamma$-fusion, inclusive photon 
analysis. 

\begin{figure}[h]
\centering
\includegraphics[width=50mm]{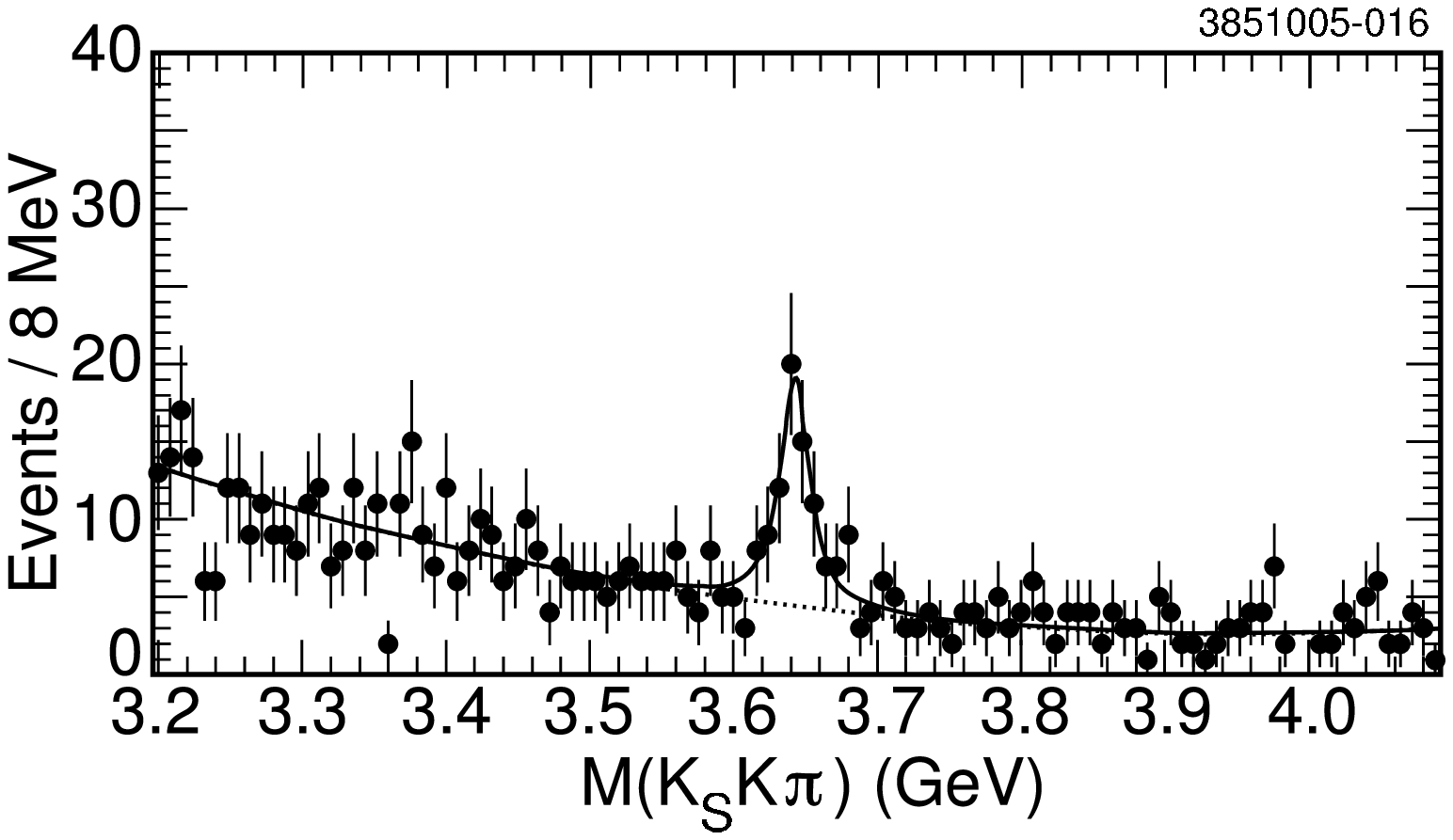}
\includegraphics[width=30mm]{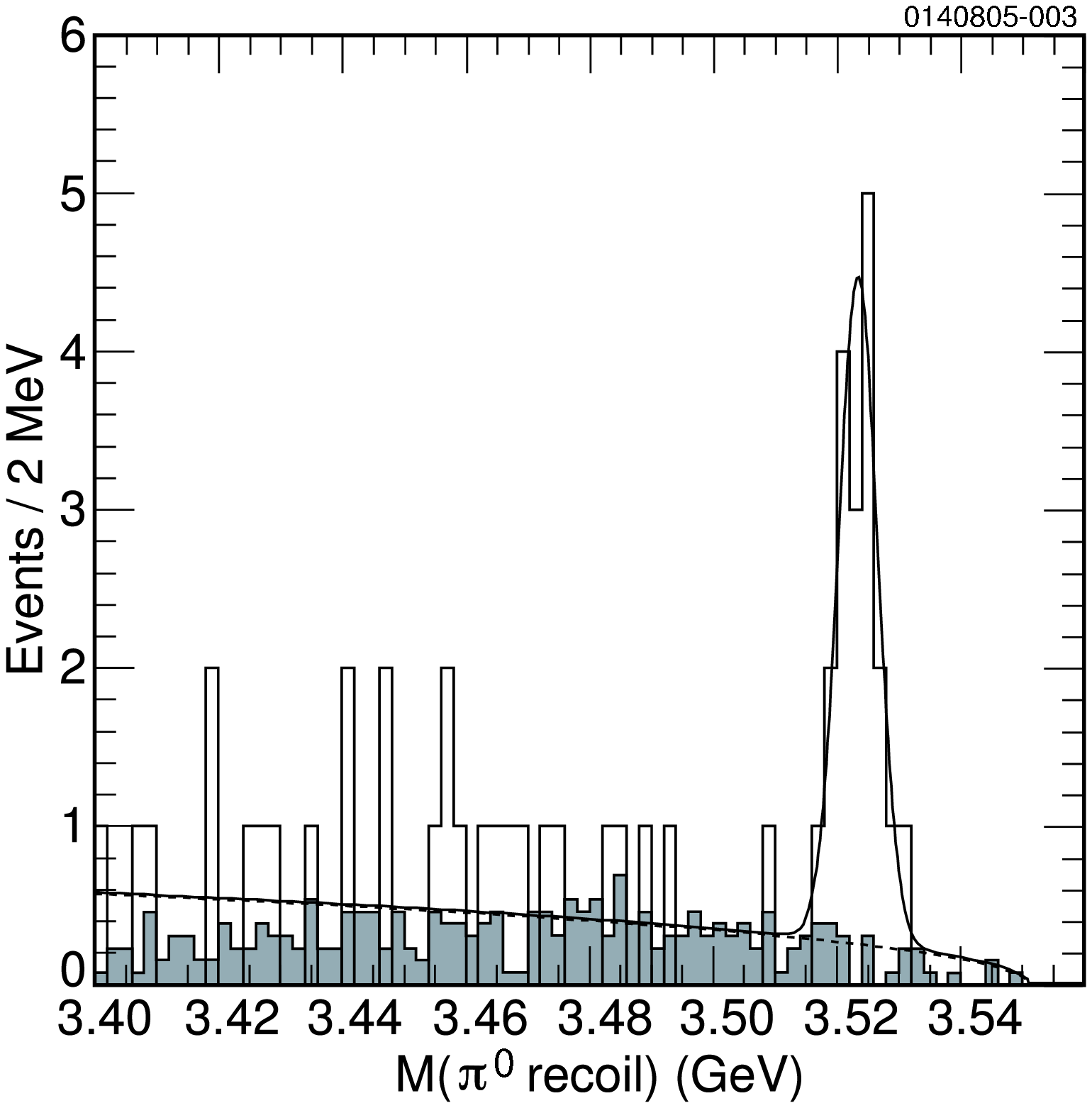}
\caption{(left) $M(K_{s}K\pi)$ from CLEO showing $\eta_{c}(2S)$.
(right) $\pi^{0}$ recoil mass spectrum from CLEO showing $h_{c}$ in
exclusive analysis of $\psi(2S) \to \pi^{0}h_{c}$, $h_{c} \to \gamma
\eta_{c}$.} \label{plot2}
\end{figure}

Finally, $\eta_{c}(2S)$ was first observed in $B$ decays by 
Belle~\cite{belleecp}. It was followed by its observation in 
$\gamma \gamma$-fusion by CLEO~\cite{cleoecp} (see Fig.~\ref{plot2} left), 
and BaBar~\cite{babarecp}. 

All available measurements of $\eta_{c}(2S)$ are summarized in 
Table~\ref{etacp}. It is obvious that the spread in mass measurements
is uncomfortably too large. The PDG 2007 weighted average value of mass
is $M(\eta_{c}(2S))$=3637$\pm$4 (MeV). This leads to the hyperfine
splitting $\Delta M_{hf}(2S)$=49$\pm$4 (MeV). The hyperfine splitting
value for $1S$ states is $\Delta M_{hf}(1S)$=117$\pm$1 (MeV). Explaining
large difference between $\Delta M_{hf}(2S)$ and $\Delta M_{hf}(1S)$
is a challenge for theorists. The width of $\eta_{c}(2S)$ is essentially 
unmeasured (PDG 2007 value is $\Gamma(\eta_{c}(2S))$=14$\pm$7 (MeV)). 
Measurement of the width is a challenge to the experimentalists.
The decay of $\eta_{c}(2S)$ is observed only in one decay channel,
$\eta_{c}(2S) \to K_{s}K\pi$.

A lot remains to be done about $\eta_{c}(2S)$. Attempts are being made 
to identify $\eta_{c}(2S)$ in the decay $\psi(2S) \to \gamma \eta_{c}(2S)$
from 54 $pb^{-1}$ CLEOc $\psi(2S)$ data. Thus, new results are expected
from CLEOc.

\begin{table}[h]
\begin{center}
\caption{Measured parameters of $\eta_{c}(2S)$ from different experiments
(PDG 2007).}
\begin{tabular}{|c|c|c|c|}
\hline
Exper. & $M(\eta_{c}(2S))$ & $\Gamma(\eta_{c}(2S))$ & Events(reaction) \\ \hline
Belle\cite{belleecp} & 4654$\pm$10 & $<$55 & 39$\pm$11 ($B \to K(K_{s}K\pi)$) \\ \hline
CLEO\cite{cleoecp} & 3643.9$\pm$3.4 & 6.3$\pm$13.0 & 61$\pm$15 ($\gamma \gamma \to K_{s}K\pi$) \\ \hline
BaBar\cite{babarecp} & 3630.8$\pm$3.5 & 17.0$\pm$8.7 & 112$\pm$24 ($\gamma \gamma \to K_{s}K\pi$) \\ \hline
BaBar\cite{babar1ecp} & 3645.0$\pm$8.4 & 22$\pm$14 & 121$\pm$27 ($e^{+}e^{-} \to J/\psi c\bar{c}$) \\ \hline
Belle\cite{belle1ecp} & 3626$\pm$8 & - & 311$\pm$42 ($e^{+}e^{-} \to J/\psi c\bar{c}$) \\ \hline
\end{tabular}
\label{etacp}
\end{center}
\end{table}

\section{Observation of $h_{c}(^{1}P_{1})$}

The observation and the measurement of the parameters of $h_{c}$
is important to determine the hyperfine splitting of P-states
$\Delta M_{hf}(1P)\equiv M(<^{3}P_{J}>)-M(^{1}P_{1})$, which is expected
to be zero from the lowest order pQCD calculations. 

Recently CLEO Collaboration has unambiguously identified $h_{c}$
in their data for 3 million $\psi(2S)$~\cite{cleo1p1}. Data have been
analyzed for the reaction $\psi(2S) \to \pi^{0}h_{c}$, $h_{c} \to
\gamma \eta_{c}$ in both inclusive and exclusive studies. In inclusive
analysis, photon energy or $\eta_{c}$ mass (recoil against $\pi^{0}\gamma$)
was constrained. In exclusive analysis seven known $\eta_{c}$ decay
channels with a total branching fraction of $\sim$10$\%$ were measured.
Results of inclusive and exclusive analysis are consistent. The final plot
of the $\pi^{0}$ recoil spectrum from exclusive analysis is shown
in Fig.~\ref{plot2} (right). 

The overall results are: \\
$M(h_{c})=(3524.4 \pm 0.6 \pm 0.4)$ MeV; \\
$\Delta M_{hf}(1P)=(+1.0 \pm 0.6 \pm 0.4)$ MeV, \\
using $M(<^{3}P_{J}>)=(3525.4 \pm 0.1)$ MeV; \\
$Br(\psi(2S) \to \pi^{0}h_{c})\times Br(h_{c}\to \gamma\eta_c)=(4.0\pm 0.8\pm 
0.7) \times 10^{-4}$; \\
Significance level of the $h_{c}$ signal is $>$6$\sigma$.

The conclusions are that a) the lowest order pQCD expectation 
$\Delta_{hf}(1P)$=0 is not strongly violated, and b) the magnitude and
the sign of $\Delta_{hf}(1P)$ is not well determined.
 
The Fermilab E835 $p\bar{p}$ annihilation experiment has also claimed
$h_{c}$ observation at $\sim$3$\sigma$ level in the reaction
$p\bar{p} \to h_{c} \to \gamma \eta_{c}$ and reported
$\Delta_{hf}(1P)$=-0.4$\pm$0.2$\pm$0.2 (MeV)~\cite{e8351p1}.

CLEOc now has new data with 24 million $\psi(2S)$ events, 
and a $h_{c}$ peaks with $\sim$250 and $\sim$1000 counts are
expected in exclusive and inclusive analysis respectively,
which will reduce the error on mass measurement more than a factor
of two.

\section{Measurements of the $\psi(2S)$ Widths}

Using $p\bar{p}$ annihilation to form charmonium $c\bar{c}$ states,
Fermilab experiment E835 achieved unprecedented precision in measuring
masses and widths of charmonium resonances. This happens due to taking
advantage of stochastically cooled antiproton beams, with FWHM energy
spreads of 0.4-0.5 MeV in the center-of-mass frame. 

\begin{figure}[h]
\centering
\includegraphics[width=60mm]{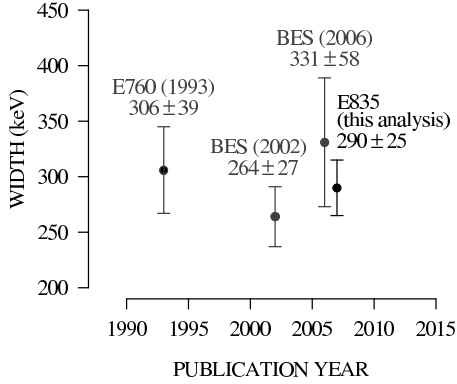}
\caption{Recent measurements of the $\psi(2S)$ widths.} \label{plot3}
\end{figure}

Recently new precision measurement of the $\psi(2S)$ total width was 
performed from excitation curves obtained in $p\bar{p}$ annihilations 
from 1.64 $pb^{-1}$ scan data in the $\psi(2S)$ region, collected by 
E835 in 2000~\cite{e835width}. The channels analyzed were 
$p\bar{p} \to e^{+}e^{-}$ and $p\bar{p} \to J/\psi X \to e^{+}e^{-} +X$. 
New technique of ``complementary scans``, based on precise beam 
revolution-frequency and orbit-length measurements was used. 
Resonance parameters were extracted from a maximum-likelihood fit to the 
excitation curves. The total width of the $\psi(2S)$ and the combination
of partial widths were measured: \\
$\Gamma_{tot}(\psi(2S))$=(290$\pm$25$\pm$4) keV, \\
$\Gamma_{e^{+}e^{-}}\Gamma_{p\bar{p}}/\Gamma_{tot}$=579$\pm$38$\pm$36 meV. \\
These represent the most precise measurements to date (Fig.~\ref{plot3}).

BES has also measured recently the $\psi(2S)$ total width using $e^{+}e^{-}$
annihilation scan data in the $\psi(2S)$ and $\psi(3770)$ regions,
collected by BES II in 2003~\cite{beswidth} (Fig.~\ref{plot3}). 
They have analyzed the channel $e^{+}e^{-} \to hadrons$. 
Resonance parameters were extracted from simultaneous fit of cross section 
curves covering energy ranges of both $\psi(2S)$ and $\psi(3770)$  
resonances: \\
$\Gamma_{tot}(\psi(2S))$=(331$\pm$58$\pm$2) keV, \\ 
$\Gamma_{ee}(\psi(2S))$=(2.330$\pm$0.036$\pm$0.110) keV.

\section{Measurements of the $J/\psi$ Widths}

Using 281 $pb^{-1}$ CLEOc $\psi(3770)$ data and looking for radiative
return events to $J/\psi$, CLEO has measured the widths of the 
$J/\psi$~\cite{cleowidth}.
They selected $\mu^{+}\mu^{-}(\gamma)$ events, each with a dimuon mass
in the region of the $J/\psi$, and counted the excess over nonresonant QED 
production. Resulting cross section is proportional to $Br_{\mu\mu} \times
\Gamma_{ee}(J/\psi)$. Assuming lepton universality and dividing by CLEO's
own measurement of $Br_{ll}$=(5.953$\pm$0.056$\pm$0.042)~\cite{cleoll},
they obtained $\Gamma_{ee}(J/\psi)$. Dividing once more by $Br_{ll}$ they
obtained $\Gamma_{tot}(J/\psi)$: \\
$Br_{\mu\mu}\times \Gamma_{ee}(J/\psi)$=(0.3384$\pm$0.0058$\pm$0.0071) keV; \\
$\Gamma_{ee}(J/\psi)$=(5.68$\pm$0.11$\pm$0.13) keV; \\
$\Gamma_{tot}(J/\psi)$=(95.5$\pm$2.4$\pm$2.4) keV. \\ 
These represent the most precise measurements to date.

\section{Measurements of the $\eta_{c}(1S)$ and $\chi_{c0,c2}(1P)$ Parameters in Two-Photon Fusion Reaction}

The masses and widths of the spin-triplet $\chi_{cJ}(1P)$ states are 
measured with high precision at Fermilab $p\bar{p}$ experiments E760/E835.
But the mass and width of the spin-singlet $\eta_{c}(1S)$ state are known
with only $\sim$1 MeV and $\sim$3 MeV precision respectively. 

\begin{figure}[h]
\centering
\includegraphics[width=80mm]{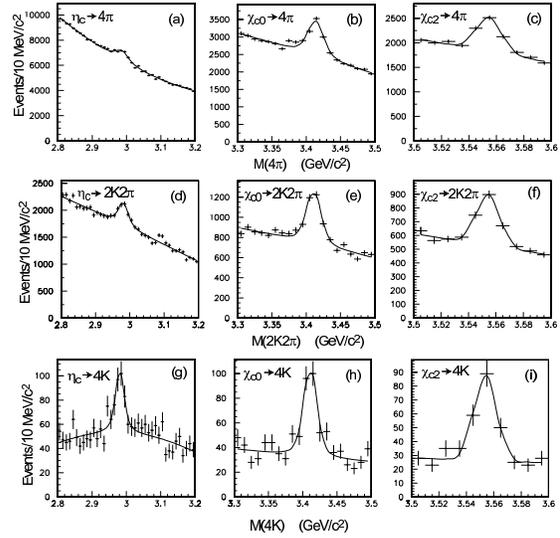}
\caption{Results of the fits to the invariant mass distributions.} \label{plot4}
\end{figure}

Using 395 $fb^{-1}$ data sample accumulated with the Belle detector,
measurements of the $\eta_{c}(1S)$ and $\chi_{c0,c2}(1P)$, produced in
two-photon collisions and decaying to four-meson final states 
(4$\pi$, 2K2$\pi$, 4K) were performed~\cite{bellegg} (Fig.~\ref{plot4}).

\subsection{Mass and width of the $\eta_{c}(1S)$ and $\chi_{c0,c2}(1P)$}

The measured values of the mass and width of $\eta_{c}(1S)$ and 
$\chi_{c0,c2}(1P)$, and the number of signal events used in analysis are 
presented in Table~\ref{bellemass}. The values of the mass and width for 
$\chi_{c0}(1P)$ and $\chi_{c2}(1P)$ are consistent within errors
with the previous high precision measurements. The precision of the 
measured mass and width of $\eta_{c}(1S)$ is comparable to other available
precision measurements.

\begin{table}[h]
\begin{center}
\caption{Mass and width measurements of the $\eta_{c}(1S)$, $\chi_{c0}(1P)$ 
and $\chi_{c2}(1P)$ from~\cite{bellegg}.}
\begin{tabular}{|c|c|c|c|}
\hline
Resonance & Mass (MeV) & Width (MeV) & N(events) \\ \hline
$\eta_{c}(1S)$ & 2986.1$\pm$1.0$\pm$2.5 & 28.1$\pm$3.2$\pm$2.2 & 7616$\pm$553 \\ \hline
$\chi_{c0}(1P)$ & 3414.2$\pm$0.5$\pm$2.3 & 10.6$\pm$1.9$\pm$2.6 & 5459$\pm$319 \\ \hline
$\chi_{c2}(1P)$ & 3555.3$\pm$0.6$\pm$2.2 & - & 2503$\pm$158 \\  
\hline
\end{tabular}
\label{bellemass}
\end{center}
\end{table}

\subsection{Two-photon widths of $\eta_{c}(1S)$ and $\chi_{c0,c2}(1P)$}

The two photon decay of the positive C-parity charmonium states in the 
lowest order is a pure QED process. The measurements of the two photon
partial widths of these states can shed light on higher order relativistic
and QCD radiative corrections.  

The values of the two photon partial widths of $\eta_{c}(1S)$ and
$\chi_{c0,c2}(1P)$, evaluated from the measurements of the 
$\Gamma_{\gamma\gamma}\times Br$ by Belle~\cite{bellegg}, are presented
in Table~\ref{belle2ppw}. They are compared to the values obtained from the
PDG 2007. The Belle value of $\Gamma_{\gamma\gamma}(\eta_{c})$ is
$\sim$2.7 times smaller ($\sim$4$\sigma$ difference) then the PDG 2007 value.
The values of $\Gamma_{\gamma\gamma}(\chi_{c0})$ and 
$\Gamma_{\gamma\gamma}(\chi_{c2})$ are consistent within errors with those
from the PDG 2007.

The ratio $R\equiv \Gamma_{\gamma\gamma}(\chi_{c2})/\Gamma_{\gamma\gamma}(\chi_{c0})$ is an interesting quantity, because it allows us to evaluate the
reliability of the first order radiative corrections, which are often very
large, by calculating $\alpha_{s}$ from them
\begin{eqnarray}
& R\equiv {\Gamma_{\gamma\gamma}(\chi_{c2}) \over \Gamma_{\gamma\gamma}(\chi_{c0})}= & \nonumber \\
& {(4|\Psi^{\prime}(0)|^{2}\alpha_{em}^{2}/m_{c}^{4}) \times(1-1.7\alpha_{s}) \over (15|\Psi^{\prime}(0)|^{2}\alpha_{em}^{2}/m_{c}^{4}) \times(1+0.06\alpha_{s})}= & \nonumber \\
& 0.267(1-1.76\alpha_{s}) & \nonumber
\end{eqnarray}

The Belle value R=0.221$\pm$0.041 leads to $\alpha_{s}$=0.098$\pm$0.085,
which is obviously underestimate of $\alpha_{s}(m_{c})$, which is known 
to be $\sim$0.3.
This makes questionable the reliability of nearly 50$\%$ first order 
correction factor for $\Gamma_{\gamma\gamma}(\chi_{c2})$.

\begin{table}[h]
\begin{center}
\caption{Two photon partial widths $\Gamma_{\gamma\gamma}$ of $\eta_{c}(1S)$, $\chi_{c0}(1P)$ and $\chi_{c2}(1P)$~\cite{bellegg}. $\Gamma_{\gamma\gamma}$ 
values are evaluated from measured $\Gamma_{\gamma\gamma}\times Br$ using 
branching fractions from PDG 2007. Results of 4$\pi$, $2K2\pi$ and 4$K$ channels are combined.}
\begin{tabular}{|c|c|c|}
\hline
Resonance & $\Gamma_{\gamma\gamma}$ (keV), Belle & $\Gamma_{\gamma\gamma}$ (keV), PDG 07 \\ \hline
$\eta_{c}(1S)$ & 2.46$\pm$0.60 & 6.7$\pm$0.9 \\ \hline
$\chi_{c0}(1P)$ & 1.98$\pm$0.24 & 2.90$\pm$0.43 \\ \hline
$\chi_{c2}(1P)$ & 0.438$\pm$0.062 & 0.539$\pm$0.050 \\ \hline
$R\equiv\Gamma_{\gamma\gamma}(\chi_{c2})/$ &  &  \\ 
$\Gamma_{\gamma\gamma}(\chi_{c0})$ & 0.221$\pm$0.041 & 0.186$\pm$0.032 \\
\hline
\end{tabular}
\label{belle2ppw}
\end{center}
\end{table}

\section{Summary}

All Charmonium states below open flavor threshold have now been firmly
identified.

The spectroscopy of spin-triplet states is now well in hand, but a lot
still needs to be done for spin-singlet states. Masses, widths, particularly
of $\eta_{c}(2S)$ and $h_{c}(1^{1}P_{1})$ need to be better determined.
Many more decay channels need to be investigated for each.

A large number of investigations, based on the world's largest sample of
$\psi(2S)$ acquired by CLEOc, are currently in progress, and results are 
expected soon. These include: \\
- Precision results for mass, width and branching fractions of 
$h_{c}(1^{1}P_{1})$; \\
- Results for many decay channels of $\eta_{c}(1S)$; \\
- Results for attempt to identify $\eta_{c}(2S)$ in radiative decay of 
$\psi(2S)$; \\
- Results of studies for $p\bar{p}$ threshold enhancement in radiative decays 
of $J/\psi$, $\psi(2S)$; \\
- Results of search for tensor glueball, $\xi(2230)$; \\
- Hadronic and radiative decays of $\psi(2S)$ and $J/\psi$; \\
- Two-body and multi-body decays of $\chi_{cJ}(1P)$ states, \\
and others. 

\bigskip

\begin{acknowledgments}
This work was supported by the U.S. Department of Energy.
\end{acknowledgments}

\bigskip 


\begin{thebibliography}{9}   

\bibitem{belleecp}   
Belle Collab. {\it Phys. Rev. Lett.} {\bf 89} (2002) 102001.

\bibitem{cleoecp}
CLEO Collab. {\it Phys. Rev. Lett.} {\bf 92} (2004) 142001.

\bibitem{babarecp}
BaBar Collab. {\it Phys. Rev. Lett.} {\bf 92} (2004) 142002.

\bibitem{babar1ecp}
BaBar Collab. {\it Phys. Rev.} {\bf D 72} (2005) 031101(R).

\bibitem{belle1ecp}
Belle Collab. {\it Phys. Rev. Lett.} {\bf 98} (2007) 082001.

\bibitem{cleo1p1}
CLEO Collab. {\it Phys. Rev. Lett.} {\bf 95} (2005) 102003;
{\it Phys. Rev.} {\bf D 72} (2005) 092004.

\bibitem{e8351p1}
E835 Collab. {\it Phys. Rev.} {\bf D 72} (2005) 032001.

\bibitem{e835width}
E835 Collab. {\it Phys. Lett.} {\bf B 654} (2007) 74-79.

\bibitem{beswidth}
BES Collab. {\it Phys. Rev. Lett.} {\bf 97} (2006) 121801.

\bibitem{cleowidth}
CLEO Collab. {\it Phys. Rev.} {\bf D 73} (2006) 051103(R).

\bibitem{cleoll}
CLEO Collab. {\it Phys. Rev.} {\bf D 71} (2005) 111103(R).

\bibitem{bellegg}
Belle Collab. {\it hep-ex/0706.3955}.

\end{thebibliography}
\end{document}